\documentclass[10pt]{article}
 \UseRawInputEncoding 
\usepackage[table]{xcolor}
\usepackage{amsmath}
\usepackage{float}
\usepackage{graphicx}
\usepackage{tabularx}
\usepackage[letterpaper, top = 1.0in, bottom = 1.0in, right=0.75in, left = 0.75in]{geometry}
\usepackage{amssymb} 
\usepackage{titling}
\usepackage{titlesec}
\usepackage{times}
\usepackage{ragged2e}

\usepackage{blindtext}
\usepackage{graphics} 
\usepackage{epsfig} 
\usepackage{bm}
\usepackage{caption}
\usepackage{subcaption}
\usepackage[noadjust]{cite}

\usepackage{dsfont}
\usepackage{relsize}
\usepackage{verbatim}
\usepackage{array}
\usepackage{hyperref}
\usepackage{graphicx}
\usepackage{multirow}
\usepackage{comment}
\usepackage{amsmath}
\usepackage{cleveref}
\usepackage{amssymb}
\usepackage{cite}
\usepackage{array}
\usepackage{booktabs}
\usepackage[justification=centering]{caption}

\pretitle{ \begin{center} \Huge}

\posttitle{\end{center}}
\preauthor{\begin {center} \normalsize}

\postauthor{\end{center}}
\date{}
\titleformat{\section}{\normalsize \bfseries \scshape}{\thesection}{1em}{}
\titleformat{\subsection}{\normalsize \bfseries}{\thesubsection}{1em}{}
\title{\Huge GNSS Radio Occultation on Aerial Platforms with Commercial Off-The-Shelf Receivers}

\author{
\vspace{5mm} 
Bryan C. Chan,
Ashish Goel, 
Jonathan Kosh, 
Tyler G. R. Reid,
Corey R. Snyder, 
Paul M. Tarantino, 
\\Saraswati Soedarmadji, 
\& Widyadewi Soedarmadji
\\
\textit{Night Crew Labs, LLC}
\vspace{5mm} 
\\
Kevin Nelson \& Feiqin Xie
\\
\textit{Texas A\&M University Corpus Christi}
\vspace{5mm} 
\\
Michael Vergalla
\\
\textit{Free Flight Lab}
}

\begin{document}
\maketitle

\section*{Abstract}
In recent decades, GNSS Radio Occultation soundings have proven an invaluable input to global weather forecasting. The success of government-sponsored programs such as COSMIC is now complemented by commercial low-cost cubesat implementations. The result is access to more than 10,000 soundings per day and improved weather forecasting accuracy. This movement towards commercialization has been supported by several agencies, including the National Aeronautics and Space Administration (NASA), National Oceanic and Atmospheric Administration (NOAA) and the U.S. Air Force (USAF) with programs such as the Commercial Weather Data Pilot (CWDP). This has resulted in further interest in commercially deploying GNSS-RO on complementary platforms. Here, we examine a so far underutilized platform: the high-altitude weather balloon. Such meteorological radiosondes are deployed twice daily at over 900 locations globally and form an essential in-situ data source as a long-standing input to weather forecasting models. Adding GNSS-RO capability to existing radiosonde platforms would greatly expand capability, allowing for persistent and local area monitoring, a feature particularly useful for hurricane and other severe weather monitoring. A prohibitive barrier to entry to this inclusion is cost and complexity as GNSS-RO traditionally requires highly specialized and sensitive equipment. This paper describes a multi-year effort to develop a low-cost and scalable approach to balloon GNSS-RO based on Commercial-Off-The-Shelf (COTS) GNSS receivers. We present hardware prototypes and data processing techniques which demonstrate the technical feasibility of the approach through results from several flight testing campaigns.

\section{Introduction} \label{sec:intro}
GNSS Radio Occultation (RO) leverages GNSS satellites rising and setting on the horizon to extract refractivity in the troposphere and the ionosphere. Traditionally, highly specialized scientific instruments have been deployed on Low Earth Orbiting (LEO) platforms to collect GNSS-RO soundings where the first LEO GPS-RO experiment was performed in 1995 by the Microlab 1 satellite~\cite{Ware1996}. Subsequently, there have been numerous successful missions that utilize GNSS signals for atmospheric and ionospheric soundings~\cite{Yue2011a}. The joint U.S.-Taiwan Constellation Observing System for Meteorology, Ionosphere, and Climate (COSMIC) was launched in 2006 and represents the first dedicated LEO constellation of six satellites employing GNSS-RO techniques~\cite{Anthes2008a, Kumar2006}. The success of COSMIC demonstrated the operational value of these soundings in weather prediction, space weather monitoring, and geodesy. The second generation, COSMIC-2, was launched in 2019 with the goal of generating nearly 5,000 RO soundings per day~\cite{Schreiner2020a} compared to COSMIC-1's 1,000-2,000~\cite{Feltz2017, Yue2014a}.

Missions like COSMIC represent a highly specialized approach to GNSS-RO, creating unique scientific instruments for GNSS-based atmospheric soundings. Others have taken a New Space approach to GNSS-RO, one characterized by nanosatellites and  Commercial-Off-The-Shelf (COTS) components. In 2008, the 3U cubesat known as the Canadian Advanced Nanosatellite eXperiment-2 (CANX-2) mission was launched with a COTS GPS receiver to perform just such an experiment~\cite{Sarda2009,Kahr2011,Kahr2013a}. This showed that useful ionospheric soundings can be attained at two orders of magnitude less cost compared to COSMIC~\cite{Swab2015}. Building on this work, GNSS-RO methods for cubesat missions were expanded, and is now a large commercial enterprise producing more than 10,000 RO soundings per day~\cite{Skone2014,E.Bowler2020,Irisov2020a}. These data products are utilized in weather prediction by the U.S. National Oceanic and Atmospheric Administration (NOAA) and U.S. Air Force (USAF) along with various other global weather and research groups~\cite{Irisov2020a}. This commercial success showcases the value of such data, where there is a continued appetite for more soundings with NOAA targeting 20,000 RO’s per day~\cite{Werner2020a}, tenfold more than what was available just a short time ago.

This has resulted in further interest in commercially deploying GNSS-RO on complementary platforms. Here, we examine a so far underutilized platform for GNSS-RO: the high-altitude weather balloon. Balloon platforms were first examined for GPS-RO in 2012, showing feasibility and scientific proof of concept~\cite{Haase2012}. Here, we further explore concepts for commercialization and wide-scale deployment. Meteorological radiosondes (weather balloons) are deployed twice daily at over 900 locations globally and form an essential in-situ data source as a long-standing input to weather forecasting models. In fact, these platforms are often used as ground truth for development of new GNSS-RO techniques and for calibration~\cite{Kuo2005a}. Adding GNSS-RO capability to existing radiosonde platforms would greatly expand capability, allowing for persistent and local area monitoring, a feature particularly useful for hurricane and other severe weather monitoring. A prohibitive barrier to entry to this inclusion is cost and complexity as GNSS-RO traditionally requires highly specialized and sensitive equipment. This paper describes a multi-year effort to develop a low-cost and scalable approach to balloon GNSS-RO based on Commercial-Off-The-Shelf (COTS) GNSS receivers. 

Challenges with commercial GNSS receivers on high altitude balloon platforms were first examined by  the authors with test flights in 2011~\cite{Carroll2011}. The proof of concept for COTS-based GNSS-RO from high altitude balloons was first flown by the authors in 2018~\cite{Chan2018}. With development of novel retrieval algorithms, it was shown that quality GNSS-RO soundings can be gathered by such a system~\cite{Chan2019}. This proof of concept formed the foundation, this paper will present hardware prototypes and data processing techniques which demonstrate the technical feasibility of the approach through results from several high-altitude balloon and fixed wing aircraft flight testing campaigns.

\section{Instrumentation}
The instrumentation to support these efforts can be broadly categorized into (1) science payloads for GNSS RO and (2) mission systems support which includes mission management and data logging. This section will describe the mission management system known as the Balloon Re-programmable Integrated Computer (BRIC) as well two generations of COTS-based GNSS RO science instruments. The first generation was named the GNSS Radio Occultation and Observable Truth (GROOT) system and represents the first COTS-based system developed for this application. The evolution of GROOT is the Aircraft In-situ and Radio Occultation (AIRO) instrument for expanded capability in a smaller form factor towards that appropriate for commercialization.

\subsection{BRIC}
\begin{figure}[H]
    \centering
    \includegraphics[width=0.8\textwidth]{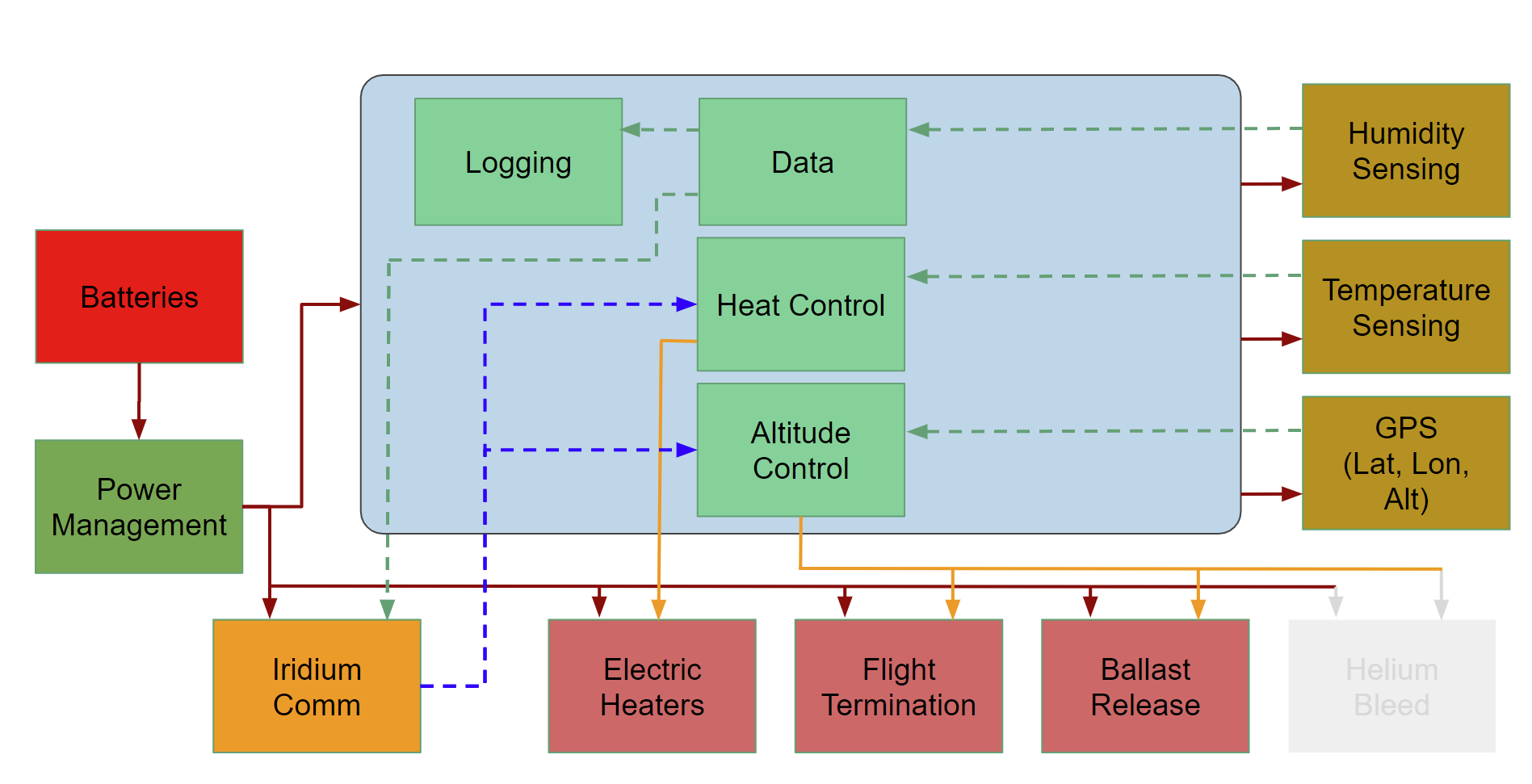}
    \caption{The BRIC flight management computer functional block diagram.}
    \label{fig:bric-block-diagram}
\end{figure}

\begin{figure}[H]
\centering
\begin{minipage}{.5\textwidth}
    \centering
    \includegraphics[width=0.95\textwidth]{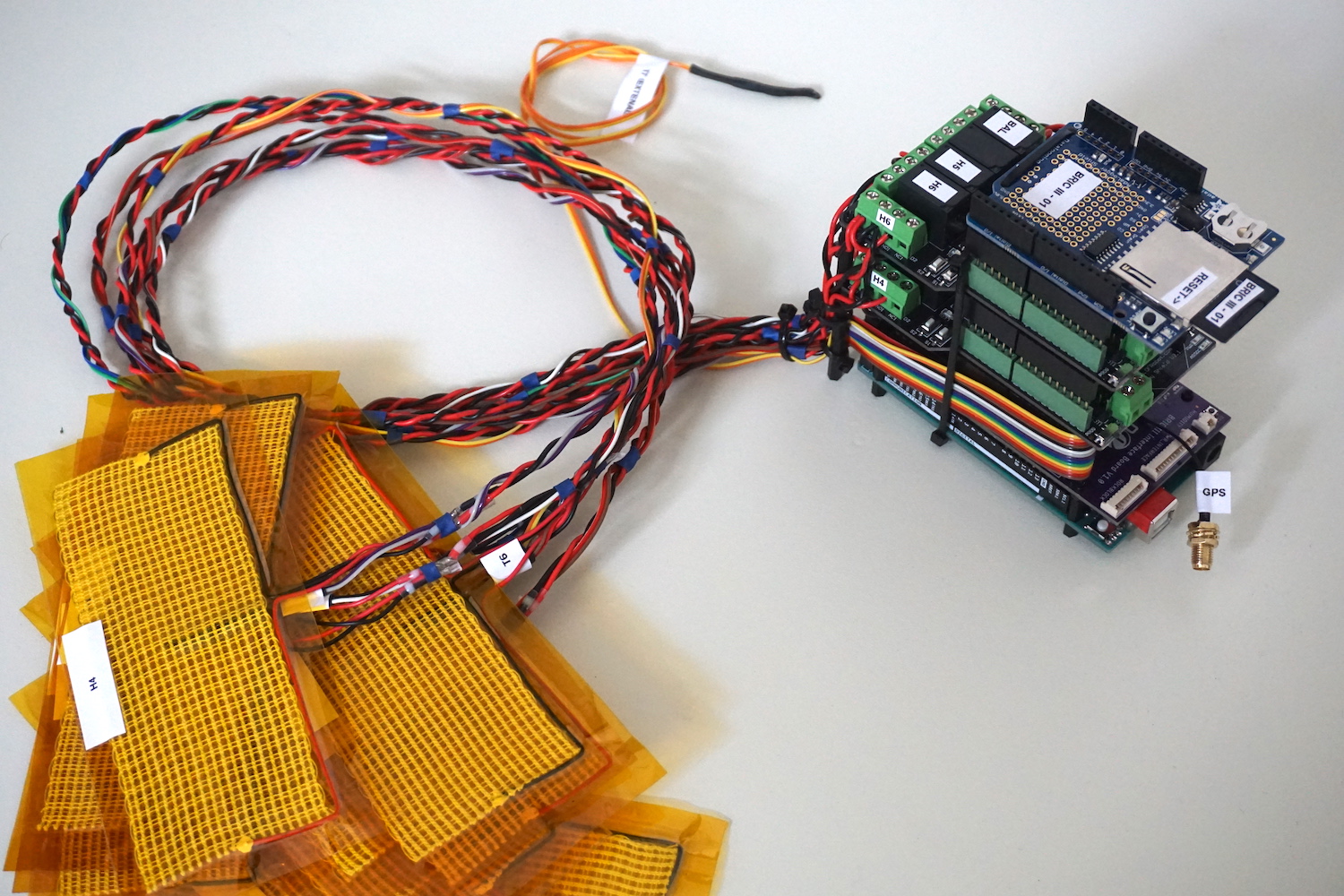}
    \caption{The BRIC flight management computer hardware. This operates the NCL HAB missions including navigation, communication, thermal management, and data logging.}
    \label{fig:bric-iii}
\end{minipage}%
\begin{minipage}{.5\textwidth}
    \centering
    \includegraphics[width=0.95\textwidth]{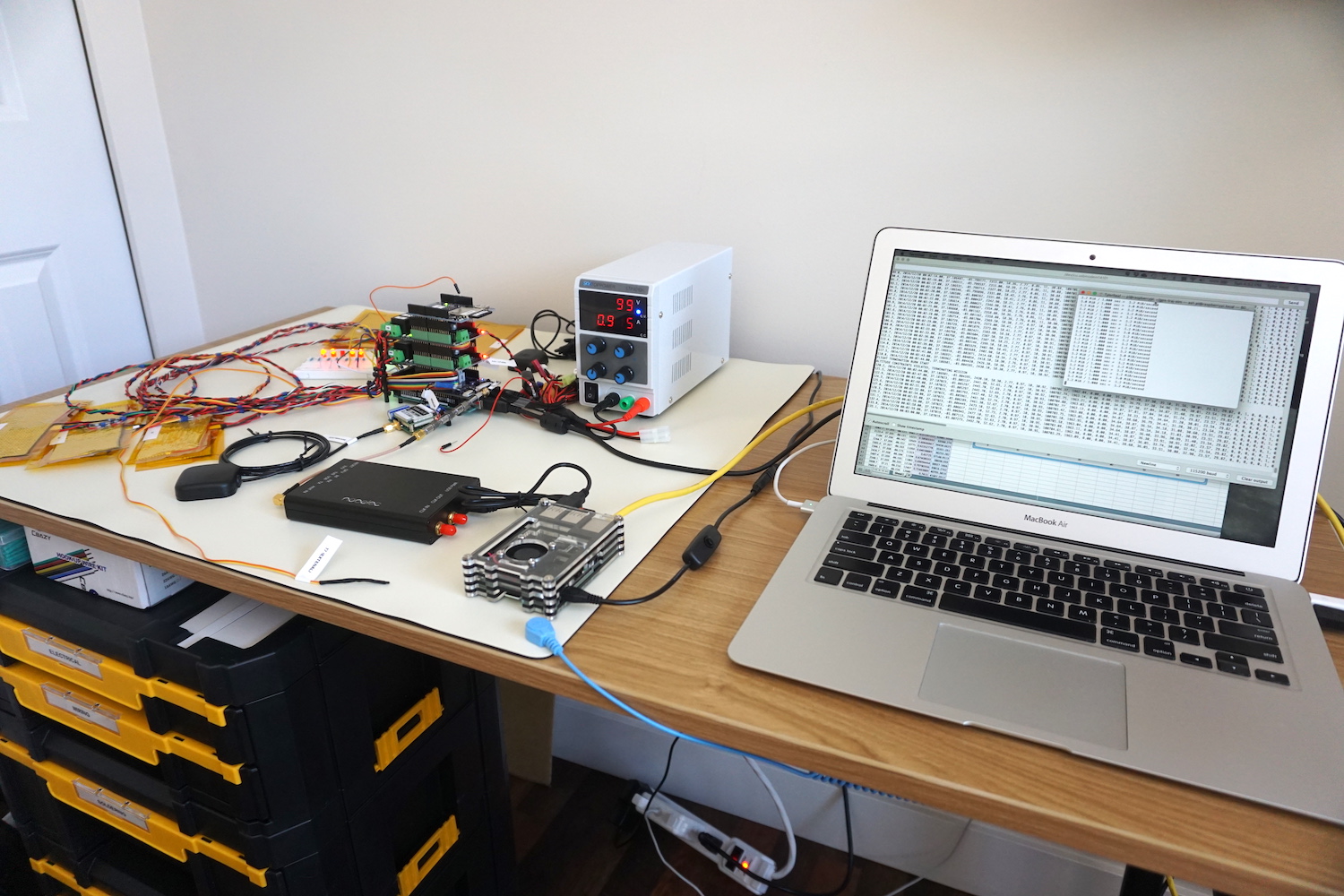}
    \caption{Hardware-in-the-loop testing of the BRIC flight computer. Shown here is a GPS simulator running through a multi-day representative flight trajectory.}
    \label{fig:bric-hil}
\end{minipage}
\end{figure}

The Balloon Re-programmable Integrated Computer (BRIC) represents the third generation of internally developed flight management computer to support Night Crew Labs (NCL) High Altitude Balloon (HAB) flights which date back to 2011~\cite{Carroll2011}. The BRIC's high-level functions are summarized in Figure \ref{fig:bric-block-diagram}. Figure \ref{fig:bric-iii} shows the hardware stack which represents a design based on an Arduino Mega 2560 for low power consumption and simplicity of implementation. Next on the stack is a custom Printed Circuit Board (PCB) which houses various sensors and interfaces. This includes barometric pressure sensors, thermistors for both environmental characterization and closed loop heating control, communication via an Iridium satellite link, and GPS for navigation. The GPS unit part of this stack was not used for radio occultation and was designed for system-level independence of the science instrument for mission reliability. 

The BRIC further provides an interface to the power management system, where current and voltage can be measured and subsystems commanded on / off as needed. Next in the stack are two COTS relay shields for control of high power load functions, including individual electric heaters, ballast release, and flight termination. Ballast release consists of an actuated valve which releases a glycol-water mixture to maintain buoyancy over several night and day cycles. The flight termination system is a nichrome wire which can be heated to cut the payload paracord for controlled descent back to the ground under parachute. On the top of the stack we have an SD card shield for data logging. These functions are summarized in Figure \ref{fig:bric-block-diagram}.

This system has been extensively tested both on the ground and in flight. A large part of the software and hardware had been tested through previous generations on more than ten successful balloon flights. The specific changes for this campaign were extensively tested on the bench. Shown in Figure \ref{fig:bric-hil} is Hardware-In-the-Loop (HIL) testing with a GPS simulator based on a software defined radio. This example is running a multi-day representative flight trajectory to qualify system behavior. 

\begin{figure} [H]
\centering
    \centering
    \includegraphics[width=0.8\textwidth]{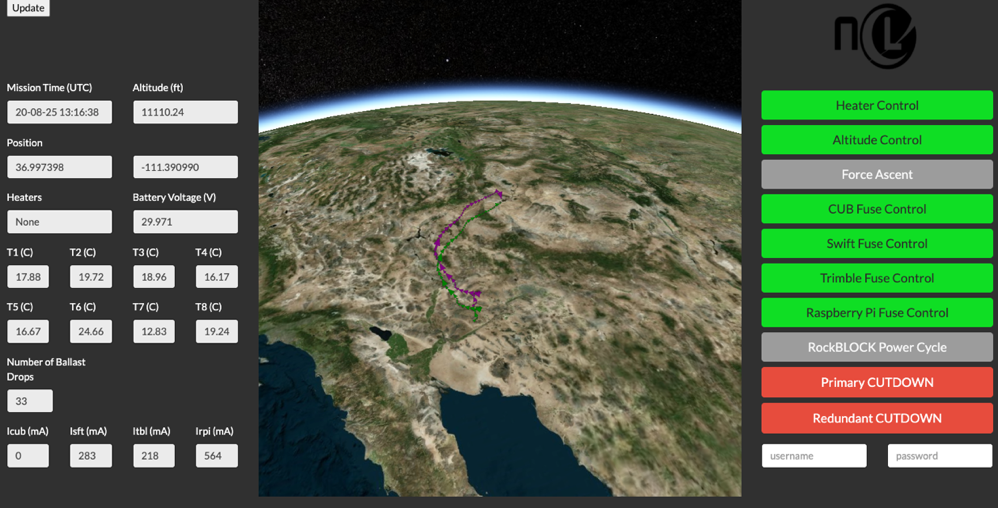}
    \caption{Custom browser-based ground station interface for balloon mission management. Iridium satellite connectivity allowed for command and control from anywhere in the world.}
    \label{fig:gs-interface}
\end{figure}

\begin{figure}[H]
 \centering
    \includegraphics[width=0.8\textwidth]{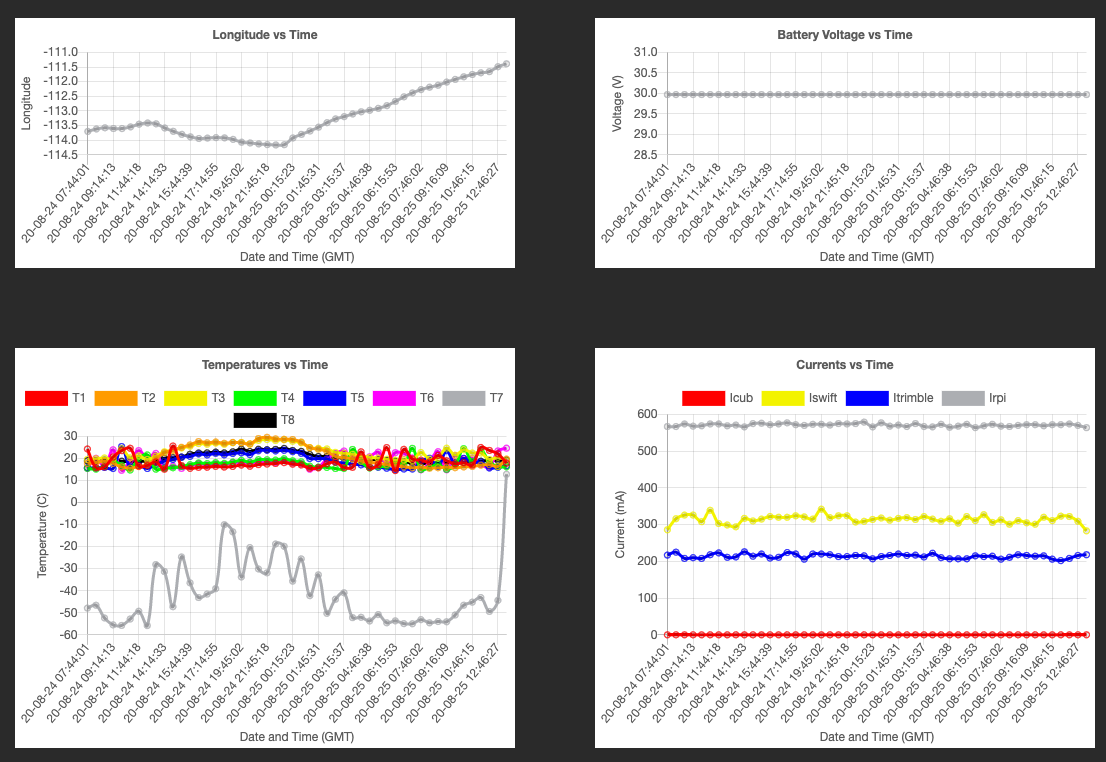}
    \caption{Ground station telemetry during flight including the position, battery voltage, internal / external temperatures, and equipment current draw. }
    \label{fig:gs-telem}
\end{figure}

The system connects via an Iridium satellite link to a custom ground station interface, where aspects of the mission can be commanded remotely via an internet browser. Functionality includes health and telemetry data as well as the ability to shut down power to individual subsystems as needed. Perhaps most importantly, this included multi-redundant flight termination capability in the event winds suddenly changed and it was deemed necessary to terminate the mission manually. Under nominal conditions, the system was intended to operate autonomously. These operations occurred in quasi-real-time as data exchange between the ground and balloon payload was set to every 2 minutes. An example of the ground station interface is shown in Figure \ref{fig:gs-interface} and data telemetry stream in Figure \ref{fig:gs-telem}.

\subsection{GROOT}
\begin{figure}[H]
    \centering
    \includegraphics[width=0.8\textwidth]{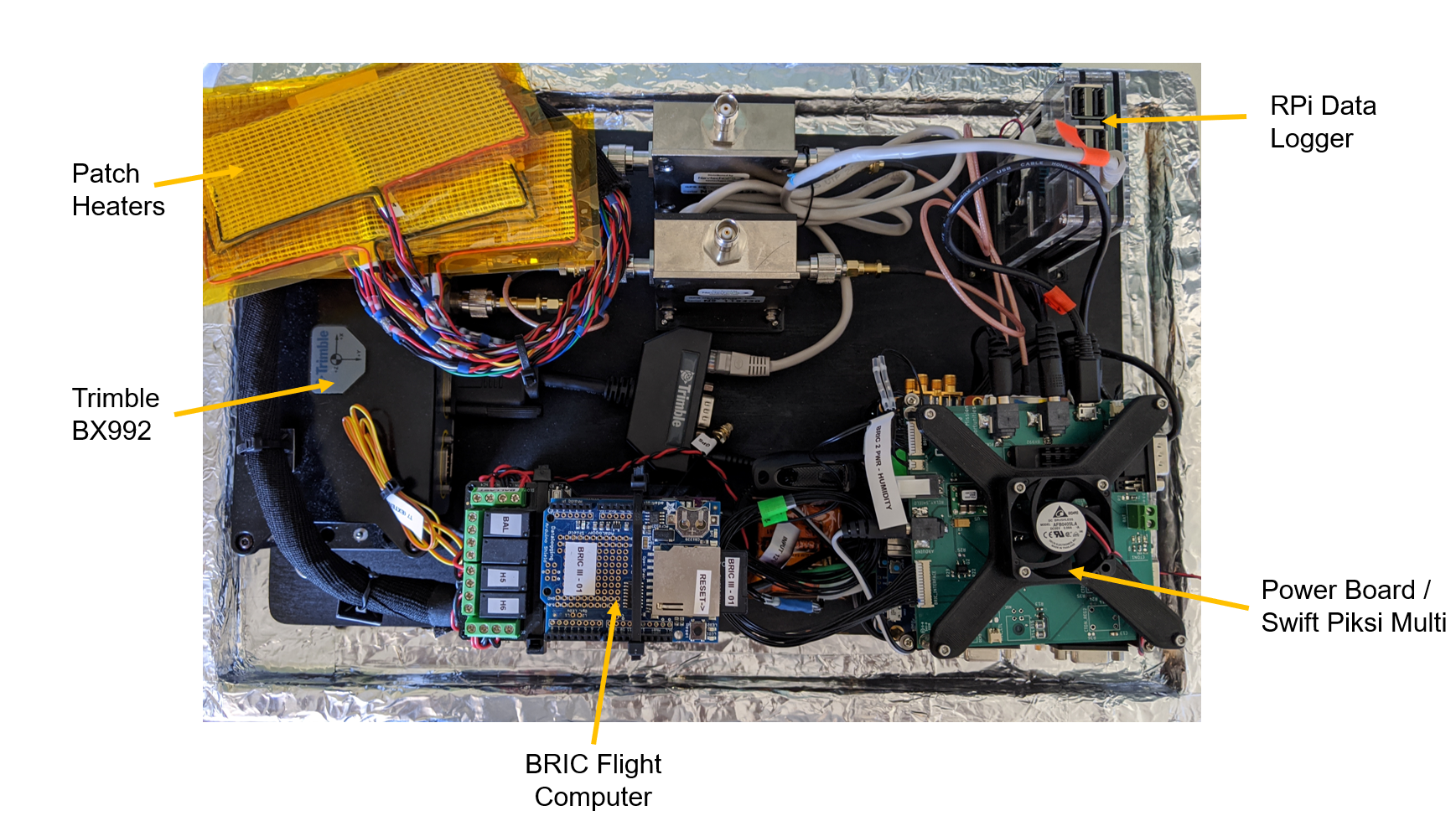}
    \caption{The GNSS Radio Occultation and Observable Truth (GROOT). A Swift Nav Piksi Multi was used for RO measurements (bottom right). A Trimble BX992 inertial navigation system (left) was used to generate precise position and velocity. Data from both were logged to a Raspberry Pi (top right).}
    \label{fig:groot}
\end{figure}

\begin{figure}[H]
    \centering
    \includegraphics[width=0.6\textwidth]{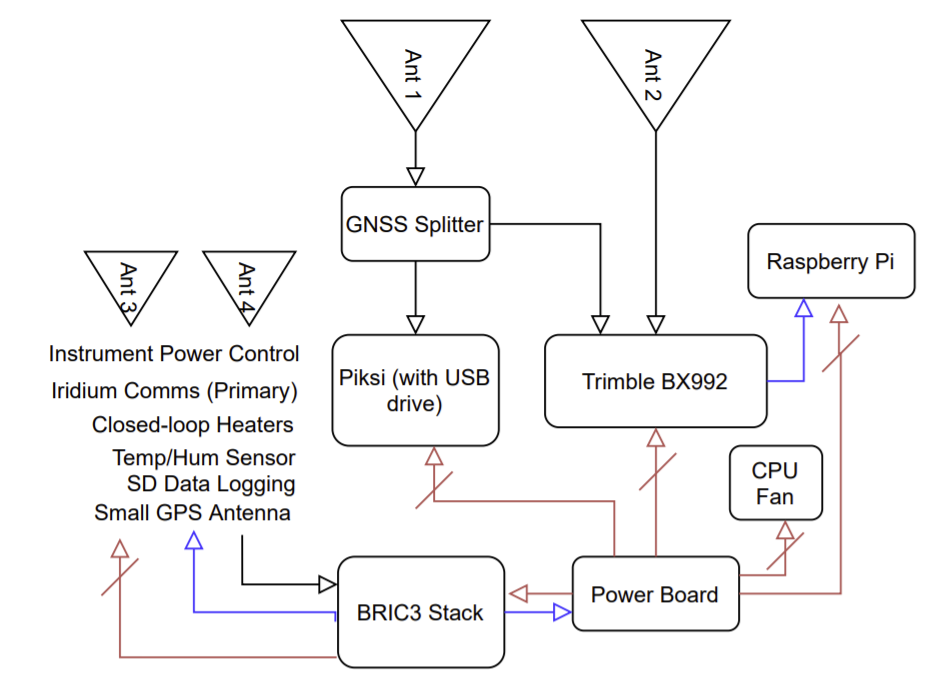}
    \caption{The GROOT RO instrument functional block diagram.}
    \label{fig:groot-block-diagram}
\end{figure}

The GNSS Radio Occultation and Observable Truth (GROOT) represents the first generation of GNSS RO science instrument based on COTS GNSS equipment by Night Crew Labs. The intent of this instrument was to collect and log the data necessary for processing and extracting GNSS RO soundings. This includes both the raw GNSS observables of the occulting satellite (e.g. carrier phase, Signal-to-Noise-Ratio (SNR), and Doppler) as well as a highly accurate position and velocity of the platform. To accomplish this, GROOT leveraged two separate GNSS devices. Raw measurements for RO where collected by a stock Piksi Multi from Swift Navigation. Precise position and velocity were computed using a dual antenna GNSS + inertial Trimble BX992 with an L-band correction service for Precise Point Positioning (PPP). Both the raw observables and precise position / velocity were logged using a Raspberry Pi computer. Figure \ref{fig:groot} shows the GROOT alongside the BRIC in a pre-flight assembly. Both GNSS units leveraged the same set of antennas and hence the need for the GNSS antenna splitters shown in the assembly. This used AV-39 aviation antennas. A functional block diagram of GROOT is shown in Figure \ref{fig:groot-block-diagram}.

\subsection{AIRO}
The evolution of GROOT is the Aircraft In-situ and Radio Occultation (AIRO) instrument. The goal of AIRO was to reduce the Cost, Size, Weight, and Power (CSWaP) of the GROOT proof of concept while maintaining RO data quality. This represents a major step towards a form factor appropriate for widespread deployment and commercialization. 
As will be discussed in Section \ref{sec:results}, COTS GNSS receivers proved highly capable in producing data of sufficient quality for RO vertical profile retrieval. Continuing this trend, the requirements for AIRO in selecting the most appropriate COTS GNSS device were that the receiver be capable of outputting raw GNSS observables at a high rate, track signals below the horizon, produce a highly accurate position / velocity solution, as well as have a small form factor and relatively low cost. As a result, the Septentrio Mosaic series was selected, being capable of outputting all necessary measurements and providing a highly accuracy position / velocity with optional L-band correction services. Satellite-based PPP corrections are necessary in HAB applications as they often fly out of sight of cellular networks. As the ultimate goal is to process RO measurements in quasi-real-time on-board the balloon, L-band corrections become a necessity. 

\begin{figure}[H]
\centering
\begin{minipage}{.5\textwidth}
    \centering
    \includegraphics[width=0.95\textwidth]{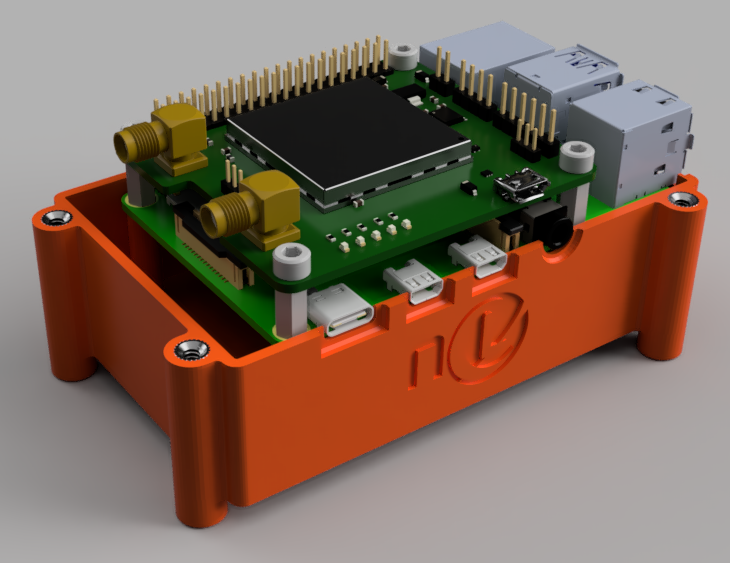}
    \caption{3D rendering of the AIRO radio occultation instrument based on the mosaicHat and Raspberry Pi 4.}
    \label{fig:airo-render}
\end{minipage}%
\begin{minipage}{.5\textwidth}
    \centering
    \includegraphics[width=0.95\textwidth]{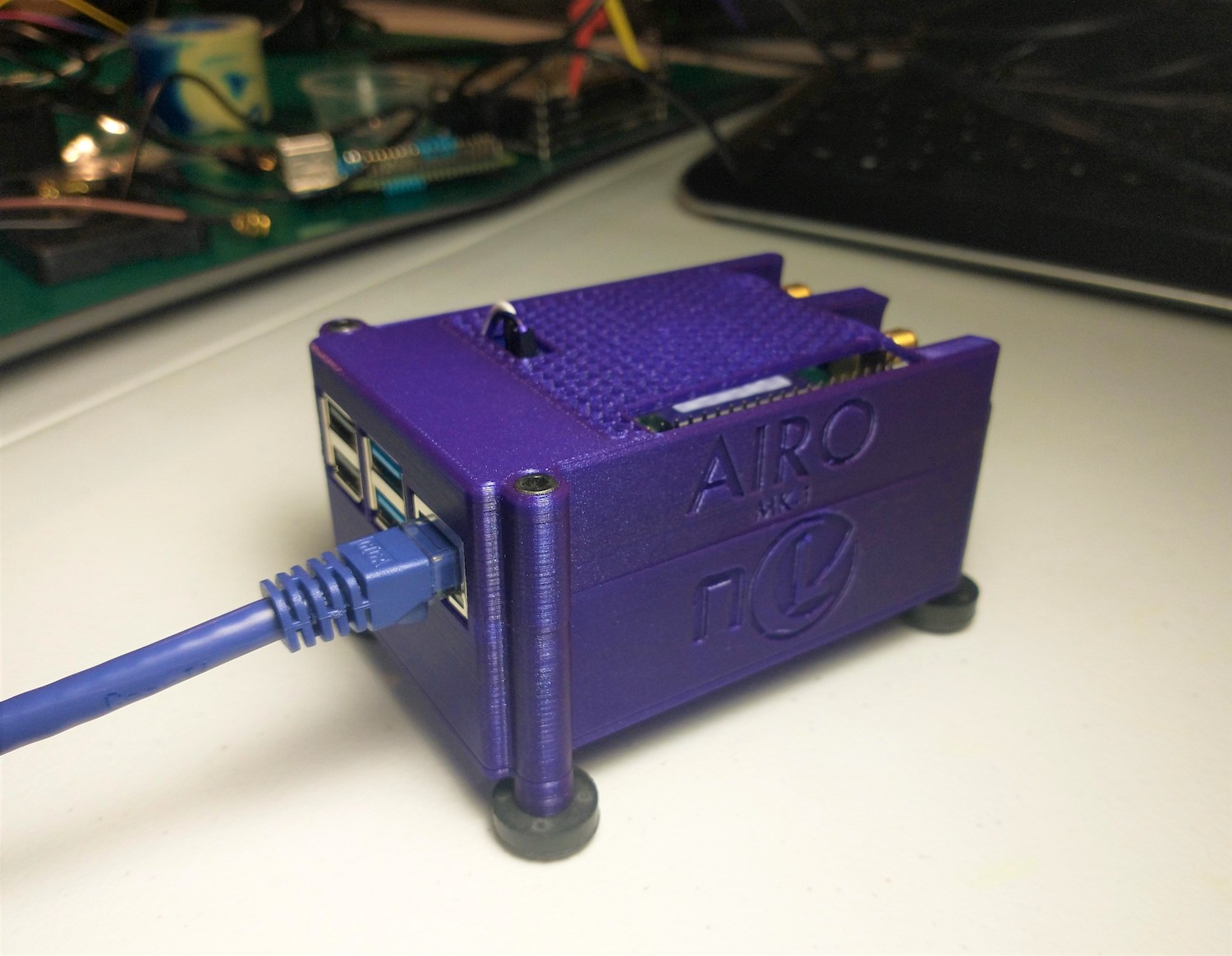}
    \caption{Completed functional prototype of the AIRO radio occultation instrument. }
    \label{fig:airo}
\end{minipage}
\end{figure}

A prototype was developed based on the mosaicHat, an open source design which interfaces the Mosaic X5 to a Raspberry Pi 4 computer~\cite{sad2020}. The design rendering is shown in Figure \ref{fig:airo-render} and the assembled functional prototype is shown in Figure \ref{fig:airo}. With correction services enabled, there is potential for this to serve as both the ground truth and RO instrument, greatly reducing the size and complexity going forward. This also reduces processing complexity, reducing the steps required for calibration between two separate GNSS systems.

\section{Flight Test Campaigns} \label{sec:campaigns}
More than ten total flight tests took place over the course of 2020 on three distinct aerial platforms. These campaigns are summarized in Table \ref{tab:summary-flight-tests}. In this section we describe seven fixed wing aircraft flights along with three high altitude balloon flights which represent more than 150 hours of flight testing with GROOT.

\begin{table}[H]
    \centering
      \caption{Summary of flight-testing campaigns.}
	\begin{tabular}{c c c}

		\toprule
        Platform &
        Number of Flights &
        Total Hours \\
		\midrule
        
        Beechcraft King Air Research Aircraft (Univ. of Wyoming) &
        7 &
        21\\
        
        World View Long Duration Balloon & 
        2 & 
        120 \\
        
        Night Crew Labs Zero Pressure Balloon &
        1 &
        12 \\
        
		\bottomrule
	\end{tabular}
  \label{tab:summary-flight-tests}
\end{table}

\subsection{Beechcraft King Air} \label{sec:campaigns-airbus}
\begin{figure}[H]
\centering
\begin{minipage}{.5\textwidth}
    \centering
    \includegraphics[width=0.95\textwidth]{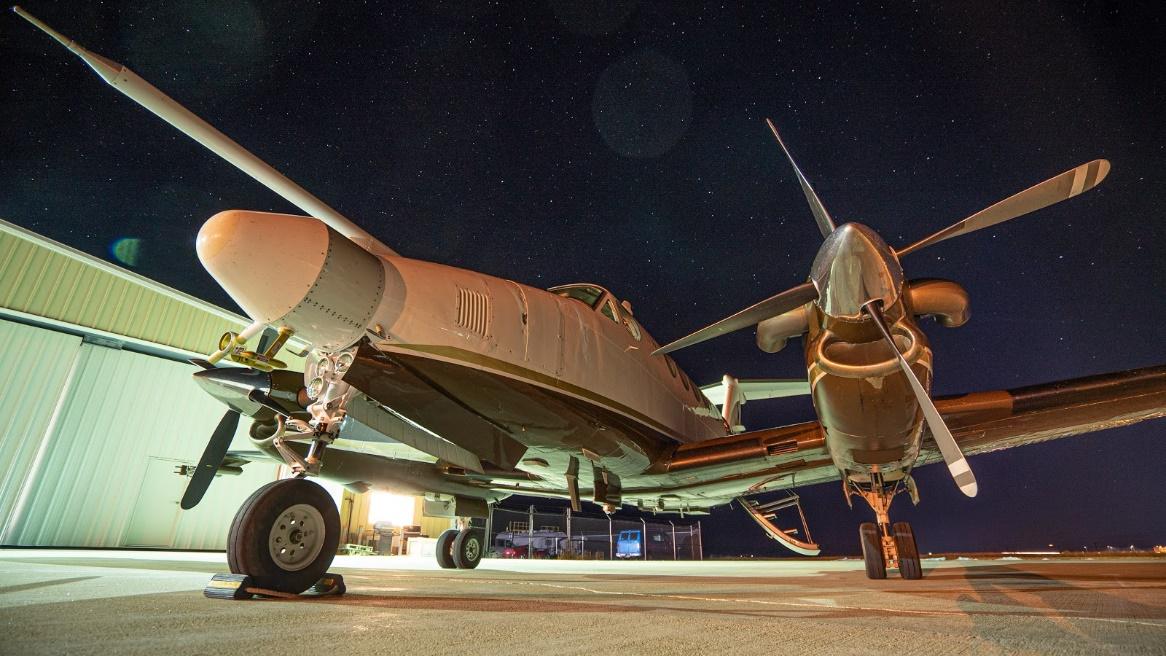}
    \caption{King Air research aircraft at the University of Wyoming.}
    \label{fig:kingair-night}
\end{minipage}%
\begin{minipage}{.5\textwidth}
    \centering
    \includegraphics[width=0.95\textwidth]{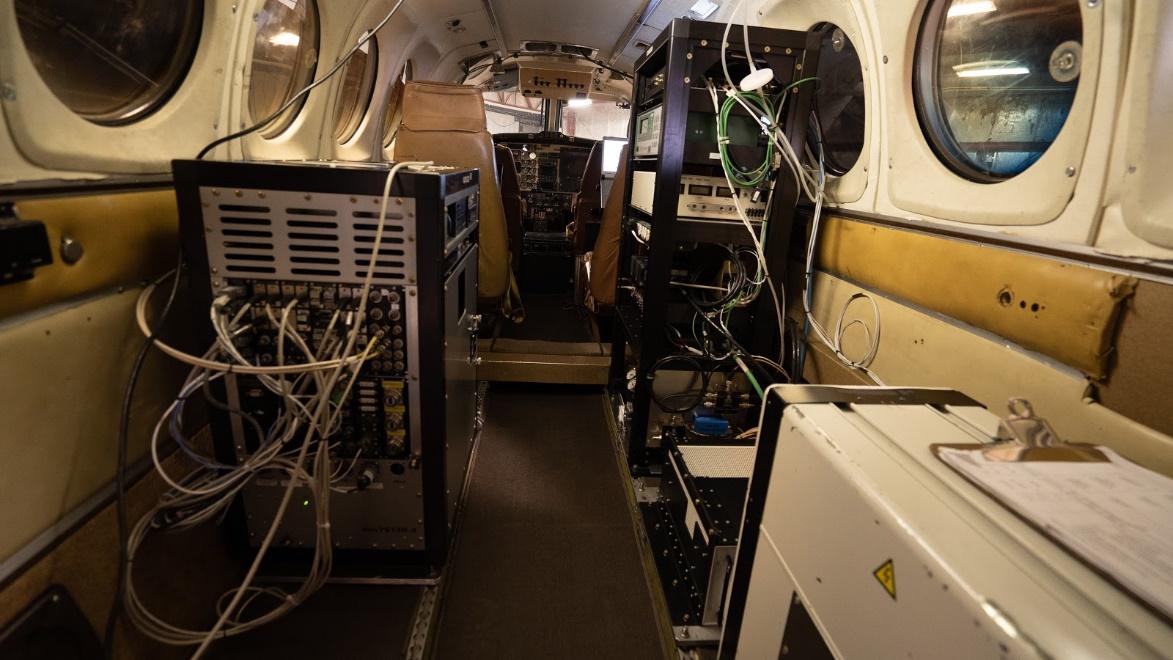}
    \caption{Instrumentation setup in the King Air research aircraft.}
    \label{fig:kingair-interior}
\end{minipage}
\end{figure}

The first of these campaigns was in February 2020 and took place on a fixed wing research aircraft platform in collaboration with Airbus Silicon Valley and the University of Wyoming. This was undertaken on a Beechcraft King Air research aircraft shown in Figure \ref{fig:kingair-night}. This entailed eight flights using the King Air over six days, collecting over 21 hours of data, where the data collection set up is shown in Figure \ref{fig:kingair-interior}. The cruising altitude was 8~km (26,000~ft).

\subsection{World View Long Duration Balloon Flight} \label{sec:campaigns-wv}
\begin{figure}[H]
\centering
\begin{minipage}{.5\textwidth}
    \centering
    \includegraphics[width=0.95\textwidth]{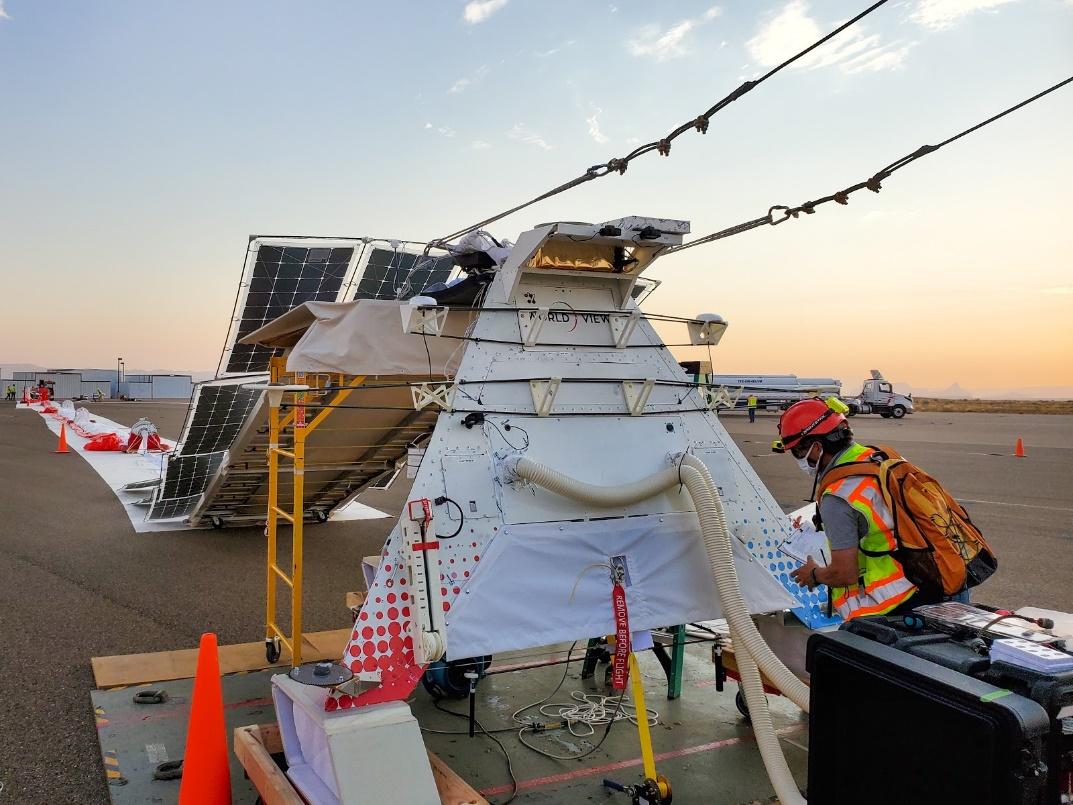}
    \caption{World View stratollite. The NCL GNSS-RO instrument was a hosted payload on the bottom face.}
    \label{fig:wv-stratollite}
\end{minipage}%
\begin{minipage}{.5\textwidth}
    \centering
    \includegraphics[width=0.95\textwidth]{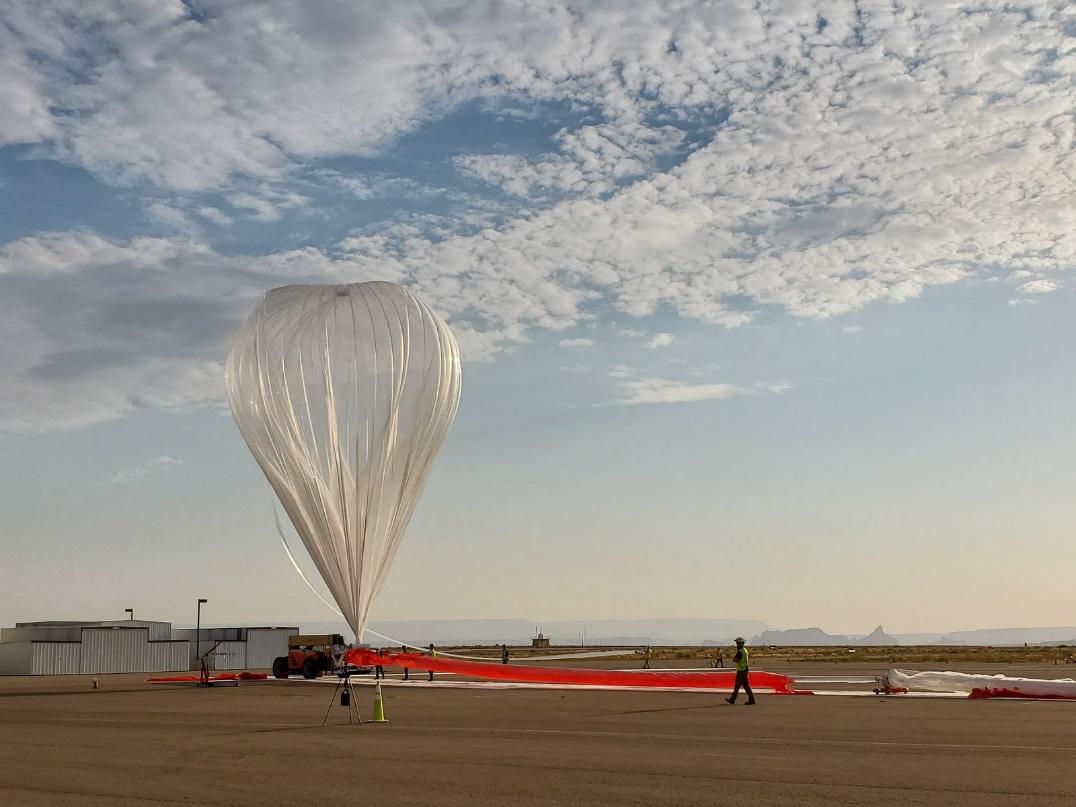}
    \caption{World View zero pressure balloon on launch day being filled with helium. }
    \label{fig:wv-launch}
\end{minipage}
\end{figure}

Following the King Air campaign was the first series of high-altitude balloon flights in the summer of 2020. These were undertaken on zero pressure balloons as a secondary hosted payload on the World View stratollite balloon bus platform shown in Figure \ref{fig:wv-stratollite}. This involved two flights out of Page, AZ which maintained 18+ km (60,000+ ft) altitude, enabling five days (120 hours) of continuous data collection. The balloon launch is shown in Figure \ref{fig:wv-launch} and the Night Crew Labs instrumentation package is shown in Figure \ref{fig:groot}. The Night Crew Labs payload included both the science instrument for RO soundings (GROOT) as well as the mission management system (BRIC), complete with tracking, satellite communication for remote command and control, as well as thermal management and data loggers. During the flight, GROOT continuously operated and collected RO data from the GPS, Galileo, and BeiDou constellations. After mission termination, the data was recovered and processed.

\subsection{NCL Zero Pressure Mission 1 (ZPM-1)} \label{sec:campaigns-ncl}

After the World View series of balloon flights, the NCL team focused on its internal Zero Pressure Mission 1 (ZPM-1) during the months of September to November. While the GROOT instrument was largely tested and validated from the World View effort, other tasks were required for preparation of an independently managed and operated HAB flight. 

\begin{figure}[H]
\centering
\begin{minipage}{.5\textwidth}
      \centering
    \includegraphics[height=0.71\textwidth]{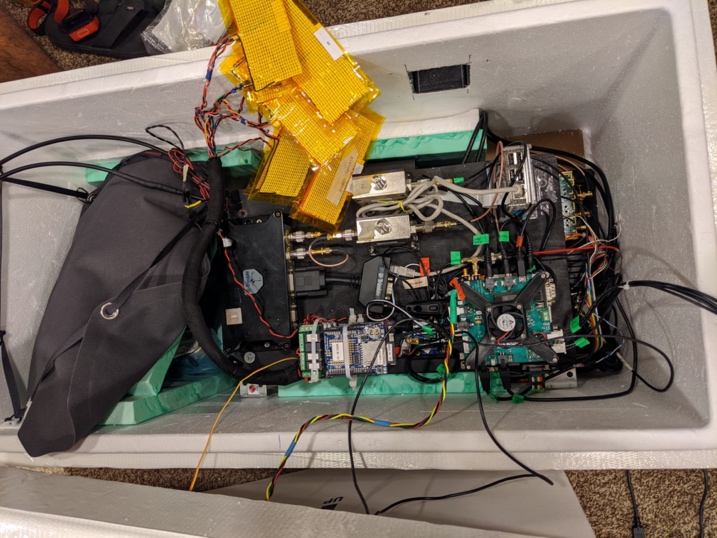}
    \caption{The NCL Zero Pressure Balloon Mission (ZPM-1) payload.}
    \label{fig:ncl-zpm-box}
\end{minipage}%
\begin{minipage}{.5\textwidth}
      \centering
    \includegraphics[height=0.71\textwidth]{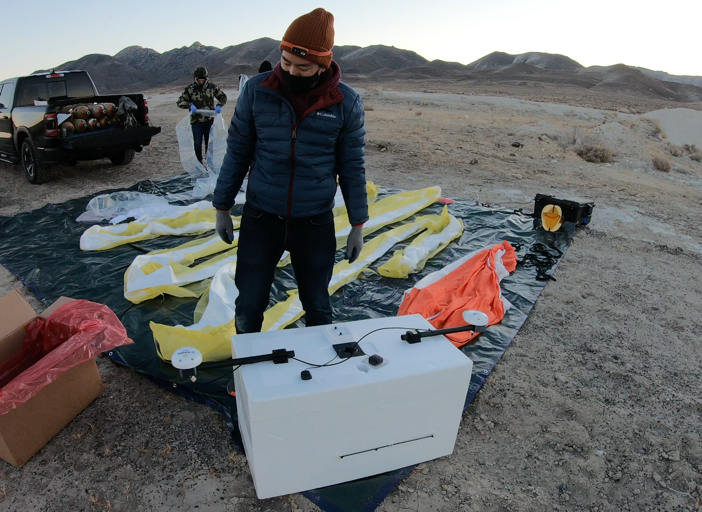}
    \caption{The NCL Zero Pressure Balloon Mission (ZPM-1) launch preparations.}
    \label{fig:ncl-zpm-prep}
\end{minipage}
\end{figure}

One major component included ballast release functionality from the BRIC, used to increase altitude during the night to combat changes in buoyancy due to natural cooling of the atmosphere. This flight also required a multi-redundant cut-down mechanism for flight termination. Due to the many components and extended duration of operation, the total payload mass was 28.5 kg (63 lbs). The payload mass necessitated using a zero-pressure balloon in lieu of a latex alternative due to its heavy-lift capacity. The payload configuration is shown in Figure \ref{fig:ncl-zpm-box}. This shows the GROOT / BRIC payloads alongside the ballast mass bladder filled with a water-glycol solution. 

\begin{figure}[H]
\centering
\begin{minipage}{0.5\textwidth}
      \centering
    \includegraphics[height=0.725\textwidth]{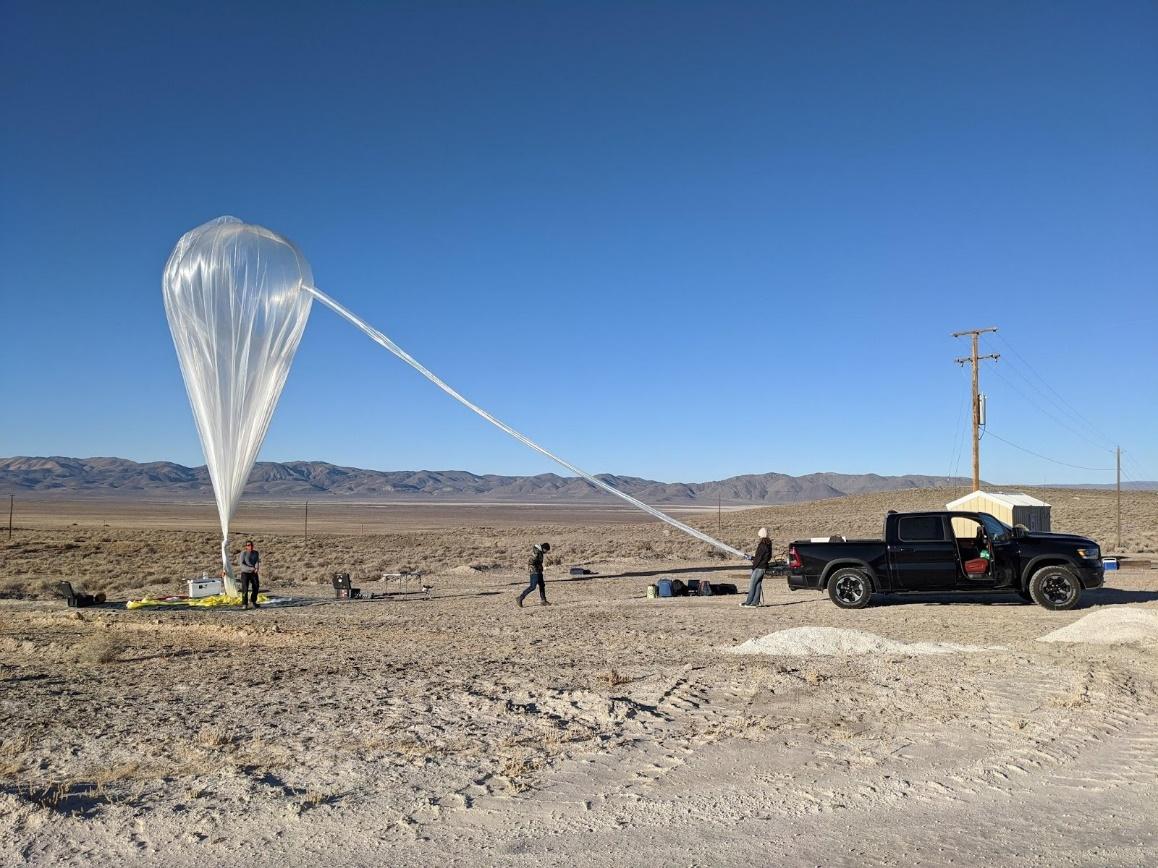}
    \caption{The NCL Zero Pressure Balloon Mission (ZPM-1) balloon filling.}
    \label{fig:ncl-zpm-launch}
\end{minipage}%
\begin{minipage}{.5\textwidth}
      \centering
    \includegraphics[height=0.725\textwidth]{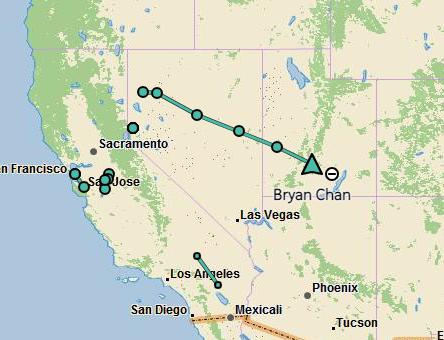}
    \caption{The NCL Zero Pressure Balloon Mission (ZPM-1) trajectory.}
    \label{fig:ncl-zpm-trajectory}
\end{minipage}
\end{figure}

Figure \ref{fig:ncl-zpm-prep} shows preparations on launch day near Empire, NV on November 28, 2020. This mission required 8.5 cubic meters (300 cu. ft.) of helium to achieve the necessary lift where Figure \ref{fig:ncl-zpm-launch} shows the filling process. The balloon ascended to a maximum altitude of 31.7 km (104,567 ft) and stayed aloft for a total of 12 hours, traveling southwest to Utah as shown in Figure \ref{fig:ncl-zpm-trajectory}. During nightfall, ZPM-1 dropped to an altitude of 17.9 km (59,000 ft) due to colder ambient temperatures at altitude. This was lower than expected, causing the balloon to drift eastwards towards the Rocky Mountains instead of the relatively flat lands of New Mexico as predicted. As such, the decision was made to end the mission after 12 hours instead of the anticipated 36 hours due to recovery considerations. The payload was recovered in southern Utah. A video of the mission can be viewed at \url{https://youtu.be/fzlcCyeYcno}.


\subsection{Future Flights}
\begin{figure}[H]
\centering
\begin{minipage}{.5\textwidth}
      \centering
    \includegraphics[height=0.725\textwidth]{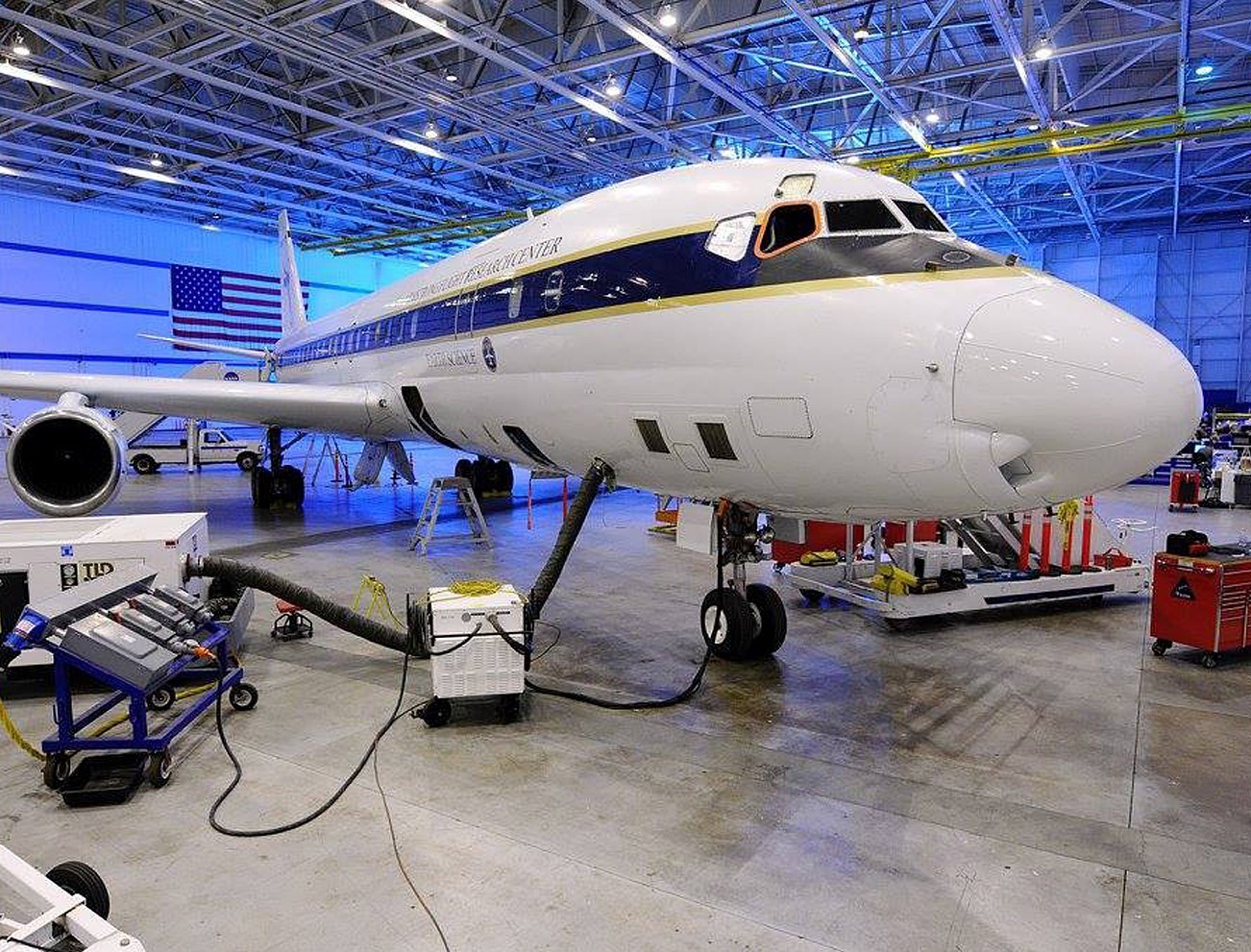}
    \caption{The NASA DC-8 research aircraft.}
    \label{fig:nasa-dc8}
\end{minipage}%
\begin{minipage}{.5\textwidth}
      \centering
    \includegraphics[height=0.725\textwidth]{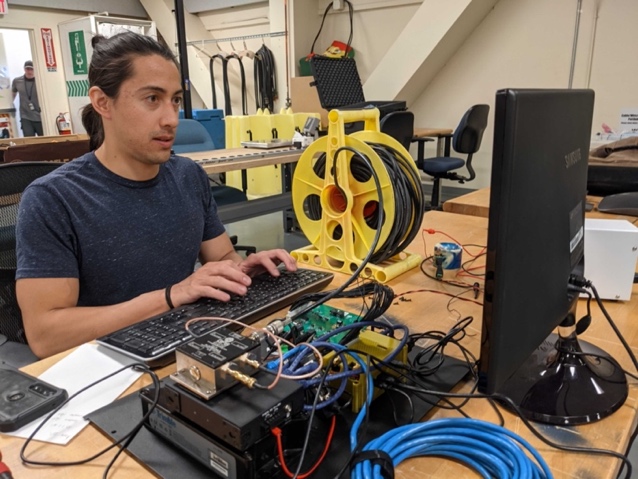}
    \caption{Testing the AIRO instrument at NASA Armstrong.}
    \label{fig:nasa-kosh}
\end{minipage}
\end{figure}

The flight testing campaigns described above conclude initial experiments with the GROOT proof of concept. Next steps will focus on AIRO. First flight tests of the AIRO instrument are scheduled to fly before the end of 2021 on the NASA DC-8 research aircraft shown at the Armstrong Flight Research Center in California in Figure \ref{fig:nasa-dc8}. Figure \ref{fig:nasa-kosh} shows testing of the AIRO instrument at NASA Armstrong which will be flown alongside GROOT for data comparison. 

\section{Results} \label{sec:results}
The focus of this paper is primarily on the hardware development and flight data collection campaigns undertaken to characterize the performance potential of COTS GNSS receivers for aerial GNSS RO. Here we give some highlights to show the significant findings, more details on the algorithms, approach, and results will be presented at a later date.

\subsection{Data Processing}
\begin{figure}[H]
    \centering
    \includegraphics[width=0.8\textwidth]{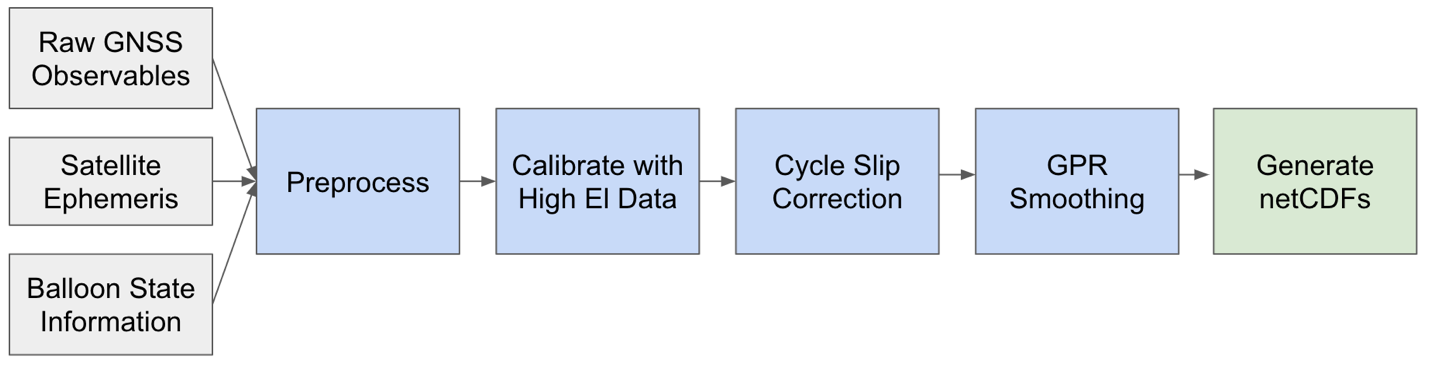}
   \caption{GNSS RO data pre-processing workflow.  }
    \label{fig:results-preprocess}
\end{figure}

After the flight data is logged, several processing steps are required to retrieve atmospheric bending angle and refractivity. The first step is to pre-process the data for later ingestion into the retrieval algorithms. Figure \ref{fig:results-preprocess} shows this workflow. The input is raw GNSS observables, satellite ephemeris data, and balloon state (position / velocity) data. Once parsed and aligned, the first step is to compute the excess phase by subtracting the line-of-sight phase from the receiver's measured phase. Step two is receiver clock calibration, where the excess phase of the occulting GNSS satellite is smoothed by subtracting the excess phase from a high-elevation GNSS satellite. Step three is cycle slip correction, where an exponential curve is fitted to the data to both smooth and remove medium to large discontinuities. Step four uses a Gaussian Process Regression (GPR) to further remove smaller discontinuities. Finally, the processed result: RO phase, RO Doppler, RO SNR, and balloon state are exported as a NetCDF file, a format convenient for downstream GNSS-RO retrieval processing.

\begin{figure}[H]
    \centering
    \includegraphics[width=0.57\textwidth]{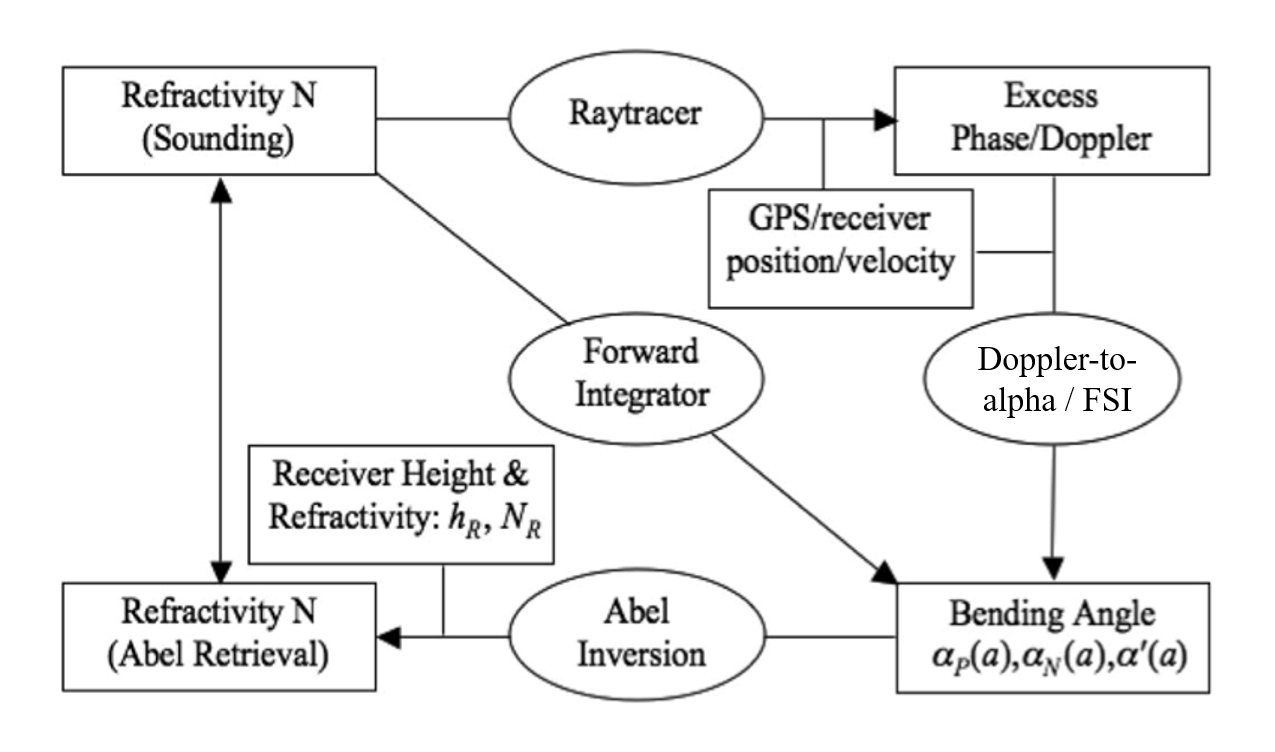}
   \caption{An end-to-end processing system for airborne / balloon-borne GNSS RO simulation, retrieval, and evaluation.   }
    \label{fig:results-algos}
\end{figure}

In order to evaluate the performance of the balloon-borne GNSS-RO retrieval algorithm, an end-to-end simulation system originally developed for aircraft-based GNSS RO was adapted for the balloon-borne RO measurements. As shown in Figure \ref{fig:results-algos}, the simulation system includes four main components: (1) a Geometrical Optics (GO) ray-tracer (ROSAP - Radio Occultation Simulator for Atmospheric Profiling), which simulates the GNSS signal as it travels through a given atmospheric refractivity model; (2) a GO retrieval module that derives the bending angle from the excess Doppler measurements (Doppler-to-Alpha)~\cite{Xie2008}, and the radio-holographic retrieval module using full-spectrum-inversion~\cite{Adhikari2016}; (3) a forward integrator that generates the bending angle profile through the forward integration of an input 1-D refractivity profile; and (4) an inverse operator to retrieve a refractivity profile via Abel inversion.

\subsection{Initial Results}
\begin{figure}[H]
    \centering
    \includegraphics[width=0.45\textwidth]{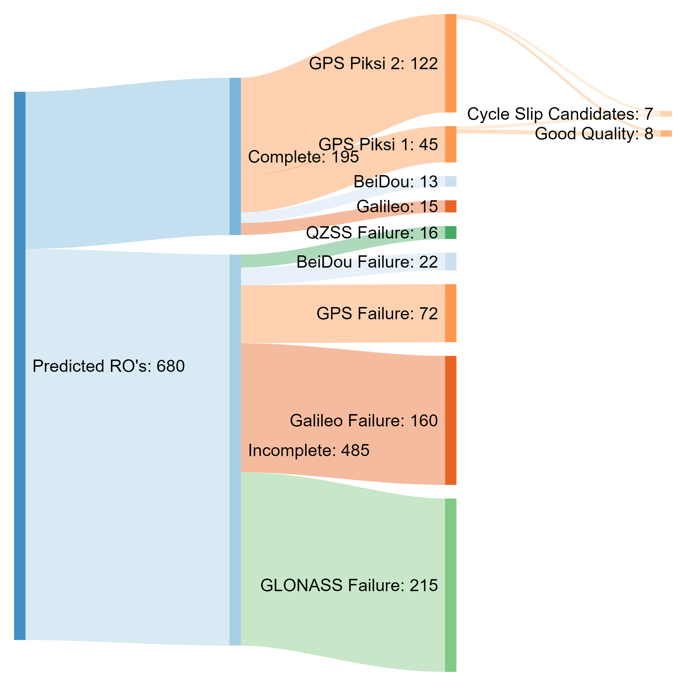}
   \caption{Sankey plot of RO data quality collected during the Wold Veiw 5-day mission. }
    \label{fig:results-sankey}
\end{figure}

In this section we show some preliminary results from the World View flight campaign. During the 120-hour flight, GROOT was exposed to 680 observable radio occultation events from GPS, Galileo, GLONASS, and BeiDou. Of these observed RO soundings, GLONASS data was not analyzed due to comparatively poor quality. RO failures from the remaining constellations were due primarily to early loss-of-lock, mainly attributed to mechanical instability and RF interference of the World View stratollite platform. After these issues, 195 RO soundings were successfully parsed. The data down selection in this process is summarized in Sankey Plot in Figure \ref{fig:results-sankey}. This flight demonstrated that RO data with a very high temporal resolution can be collected via balloon over a designated area of interest. Of these 195 RO soundings, 15 were selected for further analysis and bending angle retrieval where the median L1 excess phase of this dataset is 162~m, with a median minimum elevation angle of -4.55~degrees. 

\begin{figure}[H]
    \centering
    \includegraphics[width=0.7\textwidth]{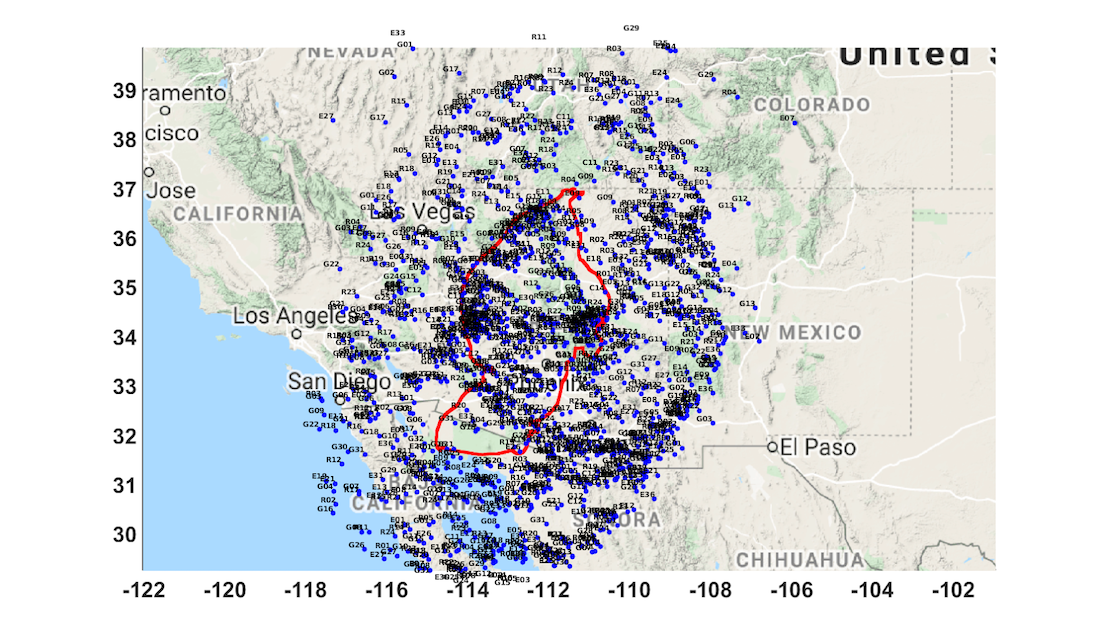}
    \caption{GNSS RO soundings recorded on a 5-day high altitude balloon flight.}
    \label{fig:wv-results-map}
\end{figure}

Figure \ref{fig:wv-results-map} shows the trajectory of the first five-day World View flight alongside the GNSS-RO profile locations. This showcases an ability to deploy such a system for persistent data collection over a local area of interest. The balloon platform showcased station-keeping ability and generally remained in Arizona airspace. Over 1,350 RO soundings were observed within an area of 900~km x 1000~km. The temporal RO sounding density, typically defined as daily RO soundings per million square miles, exceeded 340 RO soundings / 10$^6$ miles$^2$ (130 RO soundings / 10$^6$ km$^2$) for this single balloon flight, compared to approximately 50 from all operational space-based RO sources available in 2020, representing a nearly seven-fold increase.

\begin{figure}[H]
    \centering
    \includegraphics[width=0.9\textwidth]{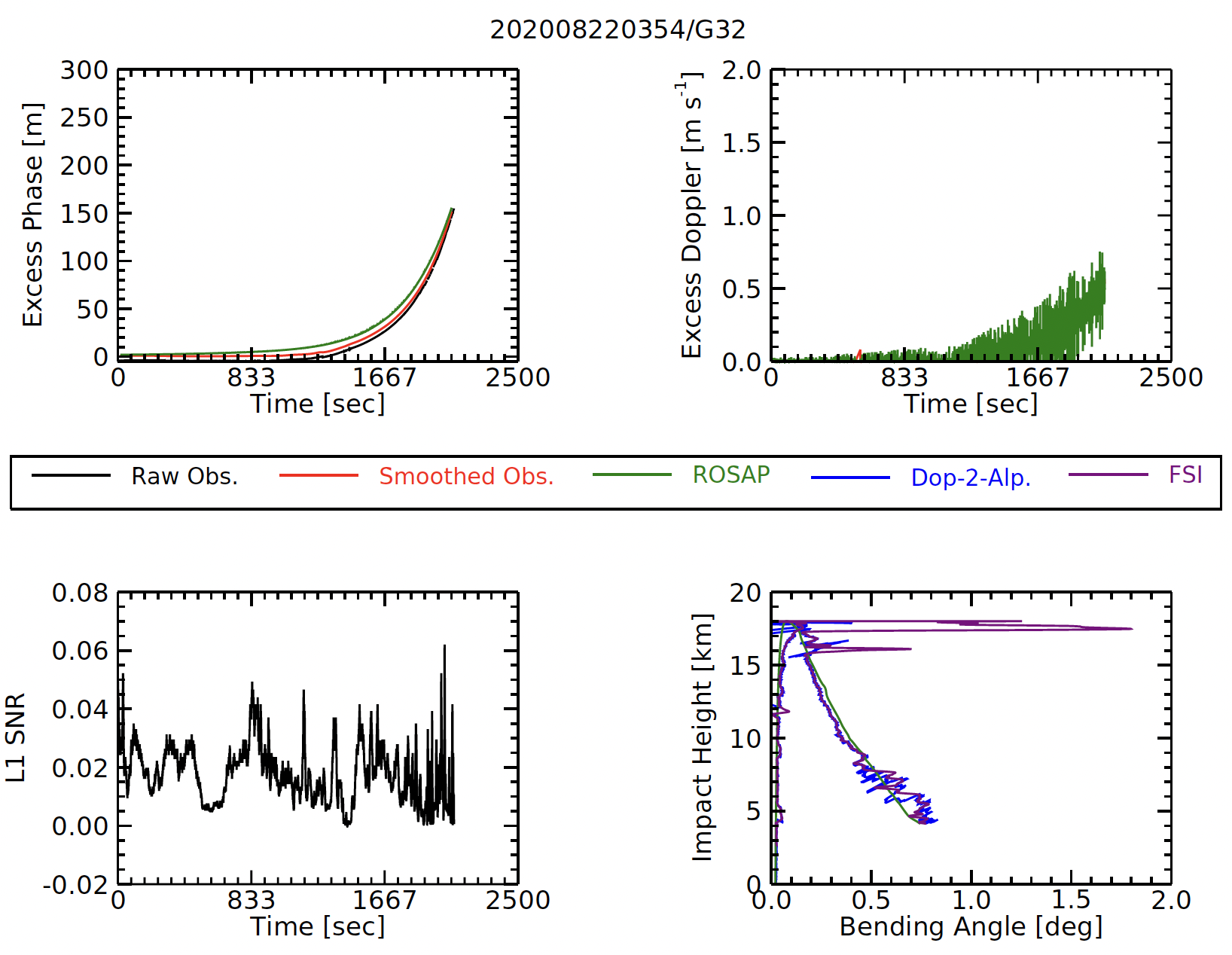}
    \caption{The G32 balloon-borne RO recorded August 22, 2020. (Top left) the excess phase and (top right) excess Doppler and simulation; (bottom left) the L1 SNR, and (bottom right) the bending angle profiles from simulation (green), GO retrieval (blue), and FSI retrieval (purple).}
    \label{fig:results-impact-height}
\end{figure}

\begin{figure}[H]
    \centering
    \includegraphics[width=0.38\textwidth]{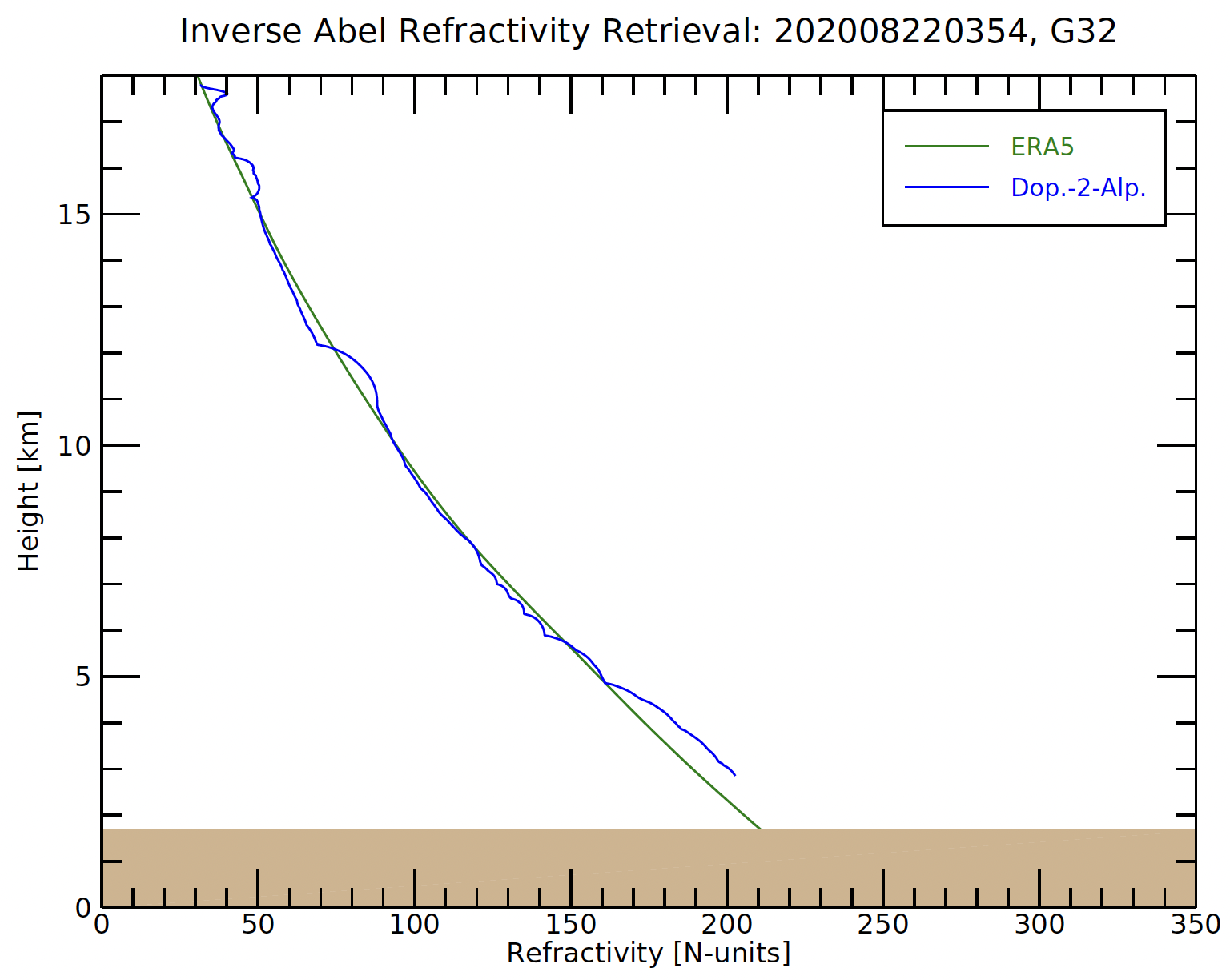}
    \caption{Example of the G32 balloon-borne RO refractivity GO retrieval (blue) compare with the close-coincident ERA5 profile.}
    \label{fig:results-refrac}
\end{figure}

The baseline for comparison was the European Centre for Medium-Range Weather Forecasts (ECMWF) fifth generation reananalysis product, ERA5~\cite{Hersbach2020}. This has a horizontal resolution of 31 km on 137 levels, from the surface up to 0.01~hPa (around 80~km). Applying the balloon RO retrieval algorithm on this data showed very good agreement between the balloon-borne RO measurements (excess phase / Doppler and bending angle) as compared to the close-coincident ERA5 profile. Figure \ref{fig:results-impact-height} shows this agreement for a specific case involving GPS PRN 32. The corresponding refractivity is shown in Figure \ref{fig:results-refrac} which shows good agreement above $\sim$~4.5~km but some biases do emerge at lower altitudes. Several other balloon-borne RO soundings showed a significant positive $N$-bias. Our end-to-end simulation study shows that such significant $N$-biases could be attributed to several factors, such as receiver position / velocity errors, the processing error in retrieval (e.g., smoothing of excess phase / Doppler, partial bending angle retrieval), and requires further investigation. It is worth noting that the Piksi receiver is capable of tracking the RO signal deep into the lower troposphere close to $\sim$~3 km above the surface.

\section{Conclusion}

This paper describes a multi-year effort to develop a low-cost and scalable approach to aerial (aircraft / balloon) GNSS RO based on COTS GNSS receivers. We present hardware prototypes and data processing techniques, which demonstrate the technical feasibility of the approach through results from several flight testing campaigns, representing more than 150 hours of data collection from both fixed wing aircraft and high altitude balloons. 

The primary science instrument described here was the GNSS RO and Observable Truth (GROOT), based on COTS-only GNSS receivers with closed-loop tracking. We demonstrated unprecedented temporal and spatial sampling density through balloon station-keeping and novel balloon-borne RO retrieval algorithms. Combined, these flight tests demonstrated that aerial platforms (e.g. balloons, aviation) are an attractive method for retrieving RO soundings over regional areas of interest for operational weather data collection.

Preliminary results are shown from a multi-day high-altitude balloon flight test over the western United States, where several hundred GNSS RO soundings were collected. Station-keeping allowed for GNSS RO soundings with high spatial and temporal resolution, resulting in a nearly seven-fold increase in RO sounding density compared to all operational space-based RO sources combined. Furthermore, this system proved capable of signal tracking down to the planetary boundary layer (2 to 3 km). Looking forward, we expect the next generation instrument known as the Aircraft In-situ and Radio Occultation (AIRO) device, to perform even better with a smaller form factor. This is slated to fly on a NASA DC-8 research aircraft in late 2021. 

The larger question surrounding these efforts is whether this method of delivering balloon-borne GNSS RO data is programmatically and commercially viable. With sufficient engineering effort, we believe that at scale, a deck-of-cards-sized GNSS RO sensor can be manufactured for less than \$1,000 with a mass of less than 1~kg. This lightweight sensor can integrate into long-duration balloon flights as a rideshare payload and deliver GNSS RO data. However, especially with the recent end of Google Loon, there are a limited number long-duration balloon flights every year which inhibits the economies-of-scale benefits. Additionally, challenges remain with integrating onto existing radiosonde platforms, as the mass requirement is closer to 100~g. 

As an adjacent technology development, we believe there is much promise in delivering airborne GNSS RO data in commercial aviation applications. A similarly sized sensor could be more easily integrated onto a commercial aircraft. This sensor could tap into an existing GNSS signal and leverage the existing radio transmission system to broadcast GNSS RO soundings to a data center in a timely fashion ($<$ 20 minutes). Additionally, tens of thousands of aircraft fly every day which immediately resolves the scaling issues of this technology.

\section{Acknowledgements}
This work was supported by the National Oceanic and Atmospheric Administration (NOAA),  National Aeronautics and Space Administration (NASA), and Airbus Silicon Valley.

\bibliographystyle{ieeetran}
\bibliography{IEEEabrv,GNSS_RO}

\section*{Biographies}
\noindent \textbf{Bryan C. Chan} is the President and CEO at Night Crew Labs. He is responsible for directing NCL's research and operational priorities, while managing the team's engineering developments. Bryan previously worked at Maxar Technologies as a program manager for geosynchronous communication satellites, and held engineering positions at SpaceX and NASA JPL. He received his B.Eng. in Aerospace Engineering from the Georgia Institute of Technology ('10), and his M.Sc. in Aeronautics and Astronautics from Stanford University ('11).
~\\

\noindent \textbf{Ashish Goel} is the
Chief Scientist of Night Crew Labs. He received his Bachelor's in Engineering Physics from the Indian Institute of Technology (Bombay) in 2009 and Ph.D. in Aeronautics and Astronautics from Stanford University in 2016. He then worked as a postdoctoral researcher in the Graduate Aerospace Laboratories
at Caltech, before joining the Jet Propulsion Laboratory, where he now serves as a Robotics Technologist in the Robotic Surface Mobility group. At NCL, he has been responsible for balloon trajectory simulation, development of GNSS RO pre-processing software, and building the flight avionics and power sub-systems.
~\\

\noindent \textbf{Jonathan Kosh} is a Software Engineer at Night Crew Labs, additionally specializing in both hardware development and data analysis. He received his B.Eng. ('11) in Aerospace Engineering from the Georgia Institute of Technology, through which he co-oped at NASA JPL. There, he helped the flight propulsion group during the assembly of the Mars Curiosity rover's powered descent stage.  Along the way, Jon realized the joy of working with small, more hands-on hardware and after graduation began work at Pololu Robotics and Electronics where he spent 8 years developing new products and building / troubleshooting microcontroller-based systems. 
~\\

\noindent \textbf{Tyler G. R. Reid} is responsible for the flight management computer at Night Crew Labs. He is also a co-founder and CTO of Xona Space Systems, a start-up focused on GNSS augmentation from Low-Earth Orbit. Tyler previously worked as a Research Engineer at Ford Motor Company in localization and mapping for self-driving cars. He was also a Software Engineer at Google and a lecturer at Stanford University where co-taught the course on GPS. Tyler received his Ph.D. ('17) and M.Sc. ('12) in Aeronautics and Astronautics from Stanford where he worked in the GPS Research Lab and his B.Eng in Mechanical Engineering from McGill (10'). He is a recipient of the RTCA Jackson Award.  

~\\
\noindent \textbf{Corey R. Snyder} is the Chief of Design at Night Crew Labs, responsible for the mechanical design of mission systems and launch operations. In addition to his current role as Director of Production and Missions at Capella Space, Corey has worked as a Project Systems Engineer at NASA Ames Research Center for the International Space Station (ISS) Space Biology Project. Corey received his M.Sc. in Aeronautical / Astronautical Engineering from the University of Colorado, Boulder ('11) and dual B.Sc. degrees in Aerospace and Mechanical Engineering from West Virginia University ('09). 

~\\
\noindent \textbf{Paul M. Tarantino} is a Night Crew Labs co-founder. He is responsible for conducting test campaigns and for development of the NCL ground station which remotely monitors and controls the payload and flight equipment. Prior to his current role as the Director of Program Management at Xona Space Systems, Paul was an Avionics, Instruments, and GNC Engineer at Blue Origin where he designed hardware-in-the-loop systems to validate rocket engine controller performance. He earned a Ph.D. ('18) in Aeronautics and Astronautics at Stanford University as a researcher in the Space Environment and Satellite Systems (SESS) Laboratory. Paul holds M.Sc.'s in both Aeronautics and Astronautics ('12) as well as Geological and Environmental Sciences ('18) from Stanford and a B.Sc. in Aerospace Engineering from University of California, San Diego ('10).

~\\
\noindent \textbf{Saraswati Soedarmadji} is a first-year undergraduate student
at Caltech who first became involved with research in high altitude balloons through a summer project
when attending high school in the Los Angeles Unified School District. In that project, she was responsible for trajectory planning and payload recovery. She then volunteered for Night Crew Labs, where she helped develop the web-based ground station software for communicating with the balloon during the mission. She also worked on visualization of radio occultation events along the balloon trajectory.

~\\
\noindent \textbf{Widyadewi Soedarmadji} is a high school student in the Los Angeles
Unified School District who aided Night Crew Labs with the task of interfacing Arduino and Raspberry Pi and in developing plotting routines for analyzing flight data. She is also a member of a local Cube Satellite club, where under the guidance of Prof. Bethany Ehlmann, she helps with the task of calculating pose and orientation estimation of a proposed satellite.

~\\
\noindent \textbf{Kevin Nelson} is a Ph.D. candidate in Coastal and Marine System Science at Texas A\&M University –- Corpus Christi, studying GNSS RO applications in complex boundary layers, specifically in tropical cyclone environments. Kevin received his M.Sc. in Atmospheric Science from the Unversity of Kansas ('15) where he studied microphysics parameterizations in models. He received his B.Sc. in Meteorology from the Florida Institute of Technology ('12) where he worked on field studies of boundary layer stability. He was awarded internships at NASA GSFC in 2018 on classification of extreme precipitation from IMERG, and at JPL in 2019 on terrestrial boundary layer height detection using GNSS RO.

~\\
\noindent \textbf{Feiqin Xie} received the B.S. degree in atmospheric science from Lanzhou University in 1998, the M.S. degree from Peking University in China in 2001, and the Ph.D. degree in atmospheric science from the University of Arizona in 2006. He then spent two-year postdoctoral training at Purdue University. From 2008 to 2012, he worked at the NASA Jet Propulsion Laboratory in Pasadena, CA. He is currently an Associate Professor in Atmospheric Science at Texas A\&M University--Corpus Christi. His current fields of interest include atmospheric remote sensing, planetary boundary layer, low clouds, and GNSS RO retrieval from various platforms (e.g., satellite, aircraft, and balloon). Dr. Xie is a member of the American Geophysical Union, American Meteorological Society, the International Radio Occultation Working Group (IROWG), and the NASA MODIS Science Team.

~\\
\noindent \textbf{Michael Vergalla} is the founder of Free Flight Research Lab, a non-profit dedicated to utilizing paraglider platforms for climate science, conservation, and pilot safety. A former Airbus Innovation Center Executive for Project Monark, he demonstrated commercial aircraft based GNSS radio occultation for the first time in 2020. With both an undergraduate and masters degree in aerospace engineering and a professional stint giving C-suite technology innovation lectures, his work focuses on finding capability boundaries for propulsion, sensing, flight operations, and manufacturing. To this end, Michael is dedicated to unlocking multiplatform flocks for advancing knowledge of Earth systems and processes.

\end{document}